# Electronic Structure of Amorphous Copper Iodide: A p-type Transparent Semiconductor

Zhaofu Zhang[1], Yuzheng Guo[2], John Robertson[1, 2*]

[1] Department of Engineering, University of Cambridge, Cambridge, CB2 1PZ, UK
[2] School of Electrical Engineering and Automation, Wuhan University, Wuhan, 430072, China
* Email: jr@eng.cam.ac.uk

**Abstract**

The atomic and electronic structure of the p-type transparent amorphous semiconductor CuI is calculated by ab-initio molecular dynamics. It is found to consist of a random tetrahedrally bonded network. The hole effective mass is found to be quite low, as in the crystal. The valence band maximum (VBM) state has a mixed I(p)-Cu($t_{2g}$)-I(p) character, and its energy is relatively insensitive to disorder. An iodine excess creates holes that move the Fermi level into the valence band, but it does not pin the Fermi level above the VBM mobility edge. Thus the Fermi level can easily enter the valence band if p-doped, similar to the behavior of electrons in In-Ga-Zn oxide semiconductors but opposite to that of electrons in a-Si:H. This suggests that amorphous CuI could make an effective p-type transparent semiconductor.

**Introduction**

Amorphous oxide semiconductors such as In-Ga-Zn oxide (IGZO) are replacing hydrogenated amorphous silicon (a-Si:H) as the main large-area semiconductors for display applications because of their higher electron mobility [1,2]. In a similar fashion, indium tin oxide (ITO) and polycrystalline ZnO are heavily used as transparent electrodes in solar cells and displays [3]. However, a weakness of this oxide electronics is that they are all n-type, whereas the p-type equivalents have much worse performance. P-type oxides must counter two problems: they have a large hole effective mass m* [4,5], and they cannot easily be doped because of defect compensation [6,7]. P-type oxides of lower m* such as the defossalites like $CuAlO_2$ or related oxides $SrCu_2O_2$ can be designed using the principles of Kawazoe *et al* [4], to have a relatively shallow Cu d core level that interacts with the oxygen p-like valence band maximum (VBM) states to lower their m* value [8-11]. However, the fundamental problem of doping in these oxides is that their VBMs are just too deep below the vacuum level so their acceptor states can be easily compensated [7].

There have been two other approaches to overcome the design of p-type oxide semiconductors, firstly to raise the VBM by using layered oxy-chalcogenides such as LaCuOS [12], or by using oxides of lower valence like SnO or $Cu_2O$ with occupied s-like lone pairs [13-16]. The problem with SnO and $Cu_2O$ is that the Sn and Cu have a reduced valence so that they have a limited range of stability. At the moment, the p-type leg of a halide perovskite is perhaps the simplest p-type transparent conductor.

Finally, for large-area electronics such as displays, there is a desire for an *amorphous* (a-) p-type semiconductor for manufacturability, to avoid the problems of grain boundaries. The defossalites have an open layer-like lattice which is unsuitable for amorphisation. The LaCuOS type compounds being layered would lose their useful properties if they lost their layers due to disordering.

There has recently been interest in zincblende (zb)-CuI, a p-type wide gap compound [17-20]. CuI has a low hole mass and reasonable hole mobility of 8 $cm^2$/V·s. Interestingly, CuI with Sn as an amorphising agent, was recently shown by Jun *et al* [21] to be an effective amorphous (a-) semiconductor with similar hole mobility to its crystalline state, and an ability to become a degenerate semiconductor by doping. Thus, a-CuI has the potential to act as a useful p-type semiconductor. However, CuI is a super-ionic conductor at higher temperatures, which might complicate matters in future.

Here we study the random network models of a-CuI, calculate their electronic structures, and explain how this p-type conductor can retain its mobility in the amorphous state, which was previously considered to be achievable only for s-like semiconductors [1,22]. Finally, we study why the network does not undergo any reconstructions which constrain its Fermi energy ($E_F$) to lie within its mobility gap, as would occur by the "Street reconstruction" [23] in amorphous silicon.

**Methods**

The structure of a-CuI is found by using ab-initio molecular dynamics (MD) on a supercell of zb-CuI. We use the VASP and CASTEP plane-wave pseudopotential codes [24,25], with a plane wave cutoff energy of 700 eV, and initially the generalized gradient approximation (GGA) as the exchange-correlation functional for electrons.

The MD process starts with 44-100 units of CuI, anneals the network at 2000 K for 4 ps, then quenches it at 10 K/ps to 300 K. The mass density of a-CuI is taken as 5.40 $g/cm^3$, 3% less than for the crystalline phase (5.57 $g/cm^3$). CuI has a relatively low melting point of 606 °C, so the anneal creates a liquid-like structure at 2000K. The quenching rate is very rapid, and will amorphise the structure, as it does experimentally for glassy metals.

In order to minimize the creation of anomalous features such as gap states or Cu clustering, it is necessary to keep a large band gap during the MD amorphisation process. An alternative Cu pseudopotential could be generated using the Opium method [26] as in Ref. [11], to create a larger band gap. Alternatively to partly correct the under-estimation of the band gap in density functional theory, here the MD is carried out using a GGA+U potential, with an on-site potential of U = 4.8 eV applied to the closed-shell Cu 3d states [27,28]. This lowers the Cu 3d states, which then repel the upper valence I p states less, thus widening the band gap. This method is used during the MD, without greatly increasing its computational cost above GGA. This process was used previously for creating a network of a-ZnO [29]. Carter [28] finds that too large a U value will destabilize the calculation for the case of $Cu_2O$.

Following MD, the amorphous CuI structure is relaxed under GGA+U to converge the force on atoms to less than 0.02 eV/Å and the energy to less than $10^{-5}$ eV/atom. The electronic partial density of states (PDOS) is then found using the more expensive screened exchange (SX) or Heyd-Scuseria-Ernerzhof (HSE) hybrid functionals [30,31]. SX or HSE mixes in a fraction of non-local Hartree-Fock exchange into the GGA and they more accurately correct the band gap error than simple GGA+U. An HF fraction of α=0.33 is found to raise the HSE06 gap of zb-CuI to the experimental value of 3.1 eV for the VASP potential.

**Results**

We firstly consider the band structure of crystalline zb-CuI. Figure 1(a) shows the band structures of zb-CuI for a VASP potential within the GGA approximation, it has a band gap of only 1.18 eV. This is much less than 3.1 eV of the experiment, but typical for many semiconductors with GGA [19]. Figure 1(b) shows the bands for GGA+U with U = 4.8 eV, giving a band gap of 1.86 eV. We note that the inclusion of U has widened the gap perturbatively by moving down the Cu 3d $\Gamma_{12}$ state, which lies at -2.0 eV below VBM in the GGA functional (Fig. 1(a)). This GGA+U gap value is wide enough to be useful for the MD calculation. Figure 1(c) shows the bands of CuI given by the HSE hybrid functional [31] with a Hartree-Fock exchange content increased from the default α=25% to α=33%, giving a band gap of 3.05 eV. Fig. 1(d) shows the bands using CASTEP with the SX functional [30], giving a band gap of 3.20 eV. Both hybrid functionals show the correct band gap close to the experimental value.

It has been noted that Cu halides are difficult materials for GW calculations [32,33]. GW based on GGA bands gives a gap of only 2.38 eV for CuBr, whereas GW of CuBr based on GGA+U gives a gap of nearer 3.07 eV, close to the experimental value. CuI has similar properties. The calculated band edge line-up to the vacuum level by different functionals is compared in Fig. 1(e). The ionisation potential was calculated by forming a CuI slab with non-polar faces. CuI atoms have a Bader charge of ±0.28.

Figure 2(a) shows the continuous random network (CRN) model of a-CuI derived from a VASP MD calculation with GGA+U (U = 4.8 eV) and subsequent relaxation also with U = 4.8 eV. The Cu atoms are copper-colored, iodide are purple, and any Sn atoms used in later models are shown as blue. zb-CuI has tetrahedral Cu and I sites, with a Cu-I bond length of 2.64 Å. In the a-CuI structure of Fig. 2, the mean Cu-I bond length is 2.7 Å.

The network of a-CuI has a tetrahedral structure with few like-atom bonds. Figure 3(a) shows the partial radial distribution functions (RDFs) of this network. In addition to the first neighbor Cu-I peak at 2.70 Å, there is a smaller Cu-Cu second neighbor peak at 2.8 Å, which also extends to larger distances. The analogous I-I distribution starts only at 3.7 Å, indicating the absence of direct I-I bonds in this network. The network size is relatively small for cost reasons because we are primarily interested in obtaining the first neighbor coordination, either tetrahedral or trigonal, and the degree of Cu clustering. The intrinsic width of the first neighbor peak of a-CuI is larger than typical amorphous solids, due to the lower rigidity of CuI ($c_{11}$ = 45.1 GPa) [17] compared to $c_{11}$ = 165.6 GPa of Si [34].

We have calculated a-CuI networks by the MD procedure from 44 to 100 CuI formula units. The latter calculations are quite time-consuming. The basic features are not found to change, including the width of the first neighbor peak. It is recognized that the smaller networks are too small for accurate RDFs, and larger ones will be developed in due course.

The main defects in this network are 3-fold coordinated sites, as was the case in a-GaAs [35,36], a-GaN [37] and a-ZnO [29]. The calculated local structure is sensitive to the density taken for a-CuI. Lower values can cause the a-CuI network to develop a locally layer-like structure with trivalent sites. These arise from bonding resembling CuI polytypes with the rhombohedral (R3c) and 'anti-litharge' (P4/nmm) tetragonal structure [38].

An unusual feature of a-CuI discussed later in Fig. S2 in the supplementary information (SI) file, is the presence of internal Cu-Cu bonds and Cu clusters in addition to the usual Cu-I bonds. These arise because CuI is a super-ionic conductor in its higher temperature phases. The Cu ion is mobile amongst a relatively stationary framework of larger iodide ions [38-40]. The incipient motion of Cu ions leads to this Cu-Cu bonding and clustering, but this aspect is inconvenient to its role as a semiconductor.

The Cu-Cu bonds can become a significant feature in the MD when carried out with potentials giving a small band gap like 1.2 eV, but they disappear if the band gap during MD exceeds ~1.8 eV. The clusters occur for smaller gap values, as hole polarons form with the Fermi level dropping into the valence band. The atomic positions of Cu atoms of a reasonable a-CuI model are highlighted in Fig. 3(b), where the I atoms are hidden, and the Cu-Cu bonds with a length cutoff of 3.0 Å are connected (longer than the averaged Cu-Cu bond length of 2.8 Å) in Fig. 3(a), showing the low evidence of direct Cu-Cu bonding in this network.

Figure 4(a) shows the calculated GGA+U density of states for this network. The Fermi level lies in midgap, so it is a semiconductor with a clean gap of 2.0 eV. Figure 4(b) shows the calculated PDOS of a-CuI by the HSE functional. This PDOS is overall quite similar to the PDOS of crystalline CuI, with a band gap of 3.0 eV, shown in Fig 4(c). The main Cu d peak of non-dispersive $\Gamma_{12}$ states lies at -2.9 eV in the valence band (VB), in between two peaks due to interacting I p-Cu d states. The lowest conduction band (CB) is mainly Cu s states. The higher CB consists of Cu-p states. An effective hole mass can be estimated from the parabolic shape of the VBM DOS using $N(E) \sim (m^*E)^{1/2}$, and is about 0.3 $m_0$, roughly as measured, and similar to the crystalline value. Thus, a-CuI has a relatively small hole m*, like the crystalline phase [17] unlike Cu oxide compounds. The small hole effective mass guarantees the considerable hole mobility, which is significant for electronics applications.

The heat formation of zb-CuI is calculated to be -0.76 eV per CuI formula unit in SX functional, close to a previous report [20], and -0.46 eV per formula unit for the amorphous phase. A similar amorphisation energy was found for a-ZnO [29].

We now discuss why the VBM of a-CuI is relatively insensitive to disorder, and its hole effective mass is similar to that of the crystal. This aspect resembles how the CBM of a-IGZO is insensitive to disorder. This effect in the metal oxides has been attributed to the cation s-like character of its CBM [2,22], but that is only a partial explanation. The CBM states of IGZO and SnO$_2$ [41,42] consist of almost entirely of cation s-states and almost none on anion states. The CBM wavefunction consists of a symmetric combination of these s-states. This causes the phase of each atomic orbital contribution to be positive, irrespective of the ring order, whether it is 6-fold (even) or 5-fold (odd). On the other hand, a ring of sp$^3$ hybrids would have a sign clash around any 5-fold ring of bonds if it tries to have any antibonding character, as a conduction band state would normally have. This is a significant disorder and would create a Hall effect sign anomaly, as in Friedman's model [43]. The sign consistency for all ring orders of s-states causes no disorder for the case of a-IGZO.

Turning to a-CuI, Jun *et al* [21] suggest that the low sensitivity of the VBM states of a-CuI to disorder is due to its iodide p orbitals at the VBM acting in a quasi s-like manner. However, this is unlikely. We know that the VBM state of crystalline CuI has a mixed I(p) – Cu(d) character [18], like that of CuAlO$_2$ [11]. It does *not* originate from just the anion. This is also true of the VBM state in the a-CuI phase, as seen by the wavefunction in Fig. 5. The a-CuI network consists of two-center bonds of Cu and I sp$^3$ hybrids, like a-GaAs. Its VBM is a near-triply degenerate states. If we consider a plane of three atoms centered on a tetrahedral Cu site, the orbital has local I(p) – Cu(t$_{2g}$) – I(p) character, with the I(p) states lying parallel, as shown in Fig. 5. *Overall*, this I–Cu–I unit could be less sensitive to disorder, in that its energy above the non-bonding Cu 3d $\Gamma_{12}$ states remains relatively constant, the orbital being more delocalized. This is seen by comparing the calculated PDOS of each phase, and with the experimental photoemission spectrum [18].

We now consider the effect of stoichiometry on the a-CuI network. Amorphous CuI is an intrinsic semiconductor with E$_F$ lying in midgap (Fig. 4). On the other hand, Fig. 6(a) shows the network of I-rich a-CuI$_{1+x}$ with a two-atom iodine excess. The DOS shows that the E$_F$ moves into the valence band. The upper valence states form a single, continuous band with the main valence band. Notably, they do not create a split-off defect band above the VBM, with E$_F$ lying in a "mini-gap" above the main valence band, thus pinning E$_F$ above the valence band mobility edge. There is some network rearrangement to accommodate the iodine excess for the a-CuI$_{1+x}$, but there is no reconstruction, similar to the Street process [23] to cause Fermi level pinning above the mobility edge. Thus there is no creation of localized trap states.

The wavefunction of this hole state of a-CuI$_{1+x}$ is shown in Fig. 6(b). It is localized mainly on orbitals on two I sites with parallel orientation. Thus, excess holes in a-CuI$_{1+x}$ behave like excess electrons in a-IGZO, where experimentally the Hall electron mobility increases continuously with increasing free-electron numbers [1,44]. This behavior is in contrast to the behavior of doped a-Si:H. There, the bonding network on n-type doped a-Si:H undergoes a reconstruction following the model of Street [23]. This reconstruction pins E$_F$ below its conduction band mobility edge, so E$_F$ remains in localized states. This is a critical process as it limits the field-effect mobility of a-Si:H thin film transistors to small values characteristic of localised states, and it is the key difference between the behavior of a-Si:H and a-IGZO [22]. Amorphous CuI follows the a-IGZO behavior, for its holes, and leads to larger hole mobility values.

A network of a-Cu$_{44}$Sn$_2$I$_{52}$ has been simulated by including the Sn amorphising agent used experimentally by Jun *et al* [21], as shown in Fig. 7(a). Sn atoms are added to the CuI supercell with its size adjusted to maintain the density. Sn is not needed to amorphise CuI in the simulation of pure a-CuI because of the high quenching rate in the MD process. Experimentally, Sn was in the +4 charge state according to XPS, so the SnI$_4$ unit satisfies the stoichiometry. The calculations find that Sn does not affect the local bonding arrangements of Cu and I much; the calculated density of states are similar to that without Sn doping, as shown in Fig. 7(b). As can be seen in the PDOS in Fig. 7(b), the E$_F$ lies in midgap, as in the case of a-Cu$_{44}$I$_{44}$ in Fig. 4(b). This is due to the ideal stoichiometry by SnI$_4$ unit. The VBM wavefunction of a-CuSnI is similar to that in Fig. 5, with localized states on local I(p) – Cu(t$_{2g}$) – I(p) character and parallel-orientated I atoms.

The presence of Cu-Cu bonds and the appearance of internal Cu clusters is a feature of CuI related to its high temperature super-ionic behavior [38-40]. The Cu ions are mobile within a framework of the larger iodide ions. There has been great discussion over the migration path of the Cu ions. Cu clustering is a precursor of the Cu migration path.

The Cu-Cu distribution in our a-CuI networks can show the formation of Cu-Cu bonds. These can be of variable length, some as short as the primary Cu-I bonds. The Cu-Cu bonds can create Cu clusters within the main Cu-I network. These bonds have the effect of broadening the Cu 3d band. As a result, larger clusters can cause E$_F$ to enter the upper valence band, thus forming a different random network.

Direct Cu-Cu bonds are generally an error of the MD process due to the under-estimation of the GGA band gap of CuI, which allows holes to form too easily in the a-CuI VBM. Cu potentials often give a GGA gap of only ~1.1 eV for zb-CuI [19]. An MD calculation with such narrow gap will allow Cu-Cu bonds to form, which are not found when other potentials are used, such as those created by the Opium method [26,11]. Alternatively, including U = 4.8 eV in the calculation increases the GGA band gap of zb-CuI to 1.86 eV, as shown in Fig. 1(b). This wider gap can inhibit the cluster formation in MD, so the Cu-Cu interactions do not form such short bonds (Fig. 3(b)).

An example of a-Cu$_{44}$I$_{44}$ network with U=0 eV in MD process is shown in Fig. S1(a). A large Cu cluster is formed within this network (Fig. S1(b)) because of the severely underestimated gap value by GGA. Bonding analysis from the RDF curve (Fig. S1(c)) shows that the average Cu-Cu bond length in this case is 2.55 Å, much shorter than in Fig. 3(a). These Cu clusters cause E$_F$ to be pinned in the valence band, rather than the correct midgap position, as verified in the PDOS in Fig. S1(d).

Figure S2 compares the effect of different U values in MD process on the final a-CuI network, including the Cu coordination number (Fig. S2(a)), the averaged Cu-Cu bond length (Fig. S2(b)), the Cu cluster size (Fig. S2(c)), and the electronic structures including band gap and Cu d orbital position (Fig. S2(d)). Clearly, with the increase of U value in MD process, the Cu-Cu averaged bond length becomes longer, and the Cu cluster size becomes smaller.

Thus Cu cluster formation can be effectively avoided by maintaining a large U during MD. With the overall consideration, we chose the U=4.8 eV in our work.

In summary, ab-initio molecular dynamics is used to calculate the atomic and electronic structure of amorphous CuI, which was recently shown to be a transparent amorphous p-type semiconductor. The amorphous phase of CuI is found to have a similar bonding to the crystalline phase. The hole mass of a-CuI is found to be relatively low like its crystalline parent, and insensitive to disorder. The valence band maximum is found to have mixed orbital character, consisting of I(p)-Cu($d_{xy}$) multi-center orbitals. The presence of hole states due to excess I or shifting the Fermi level in a transistor does not cause a reconstruction that pins $E_F$ above the VB mobility edge, so that a-CuI is potentially very useful as a p-type amorphous transparent semiconductor.

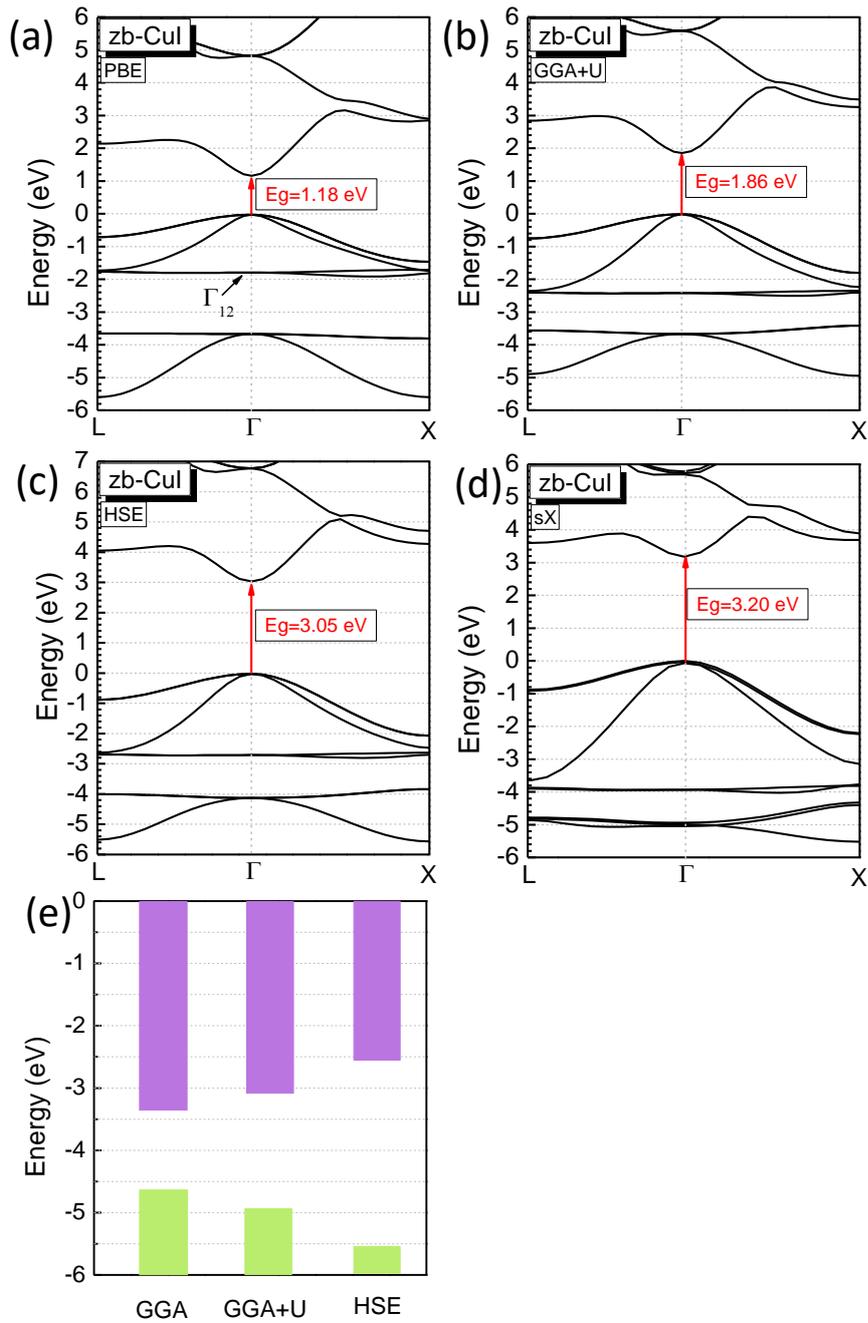

Figure 1. Band structures of zinc-blende CuI, calculated by (a) PBE, (b) GGA+U with U = 4.8 eV, (c) HSE hybrid functional with 33% Hartree-Fock non-local exchange, and (d) SX hybrid functional. (e) shows the band edge line-up of zb-CuI by different functionals, where the vacuum level is at 0 eV.

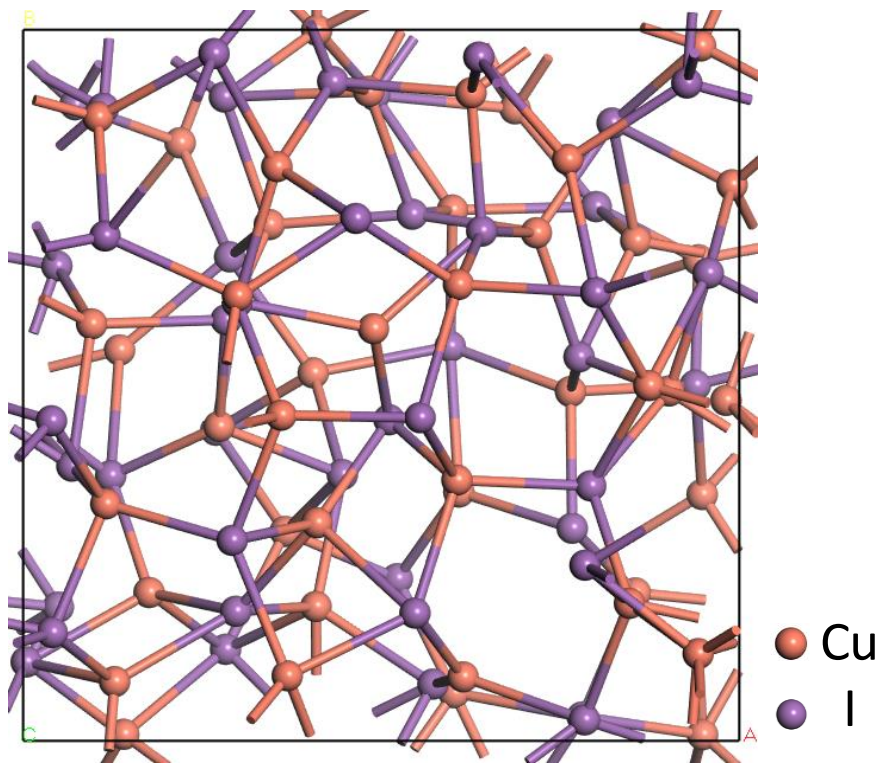

Figure 2. Random network of a-$Cu_{44}I_{44}$ created by molecular dynamics with GGA+U (U = 4.8 eV) in MD and relaxation processes.

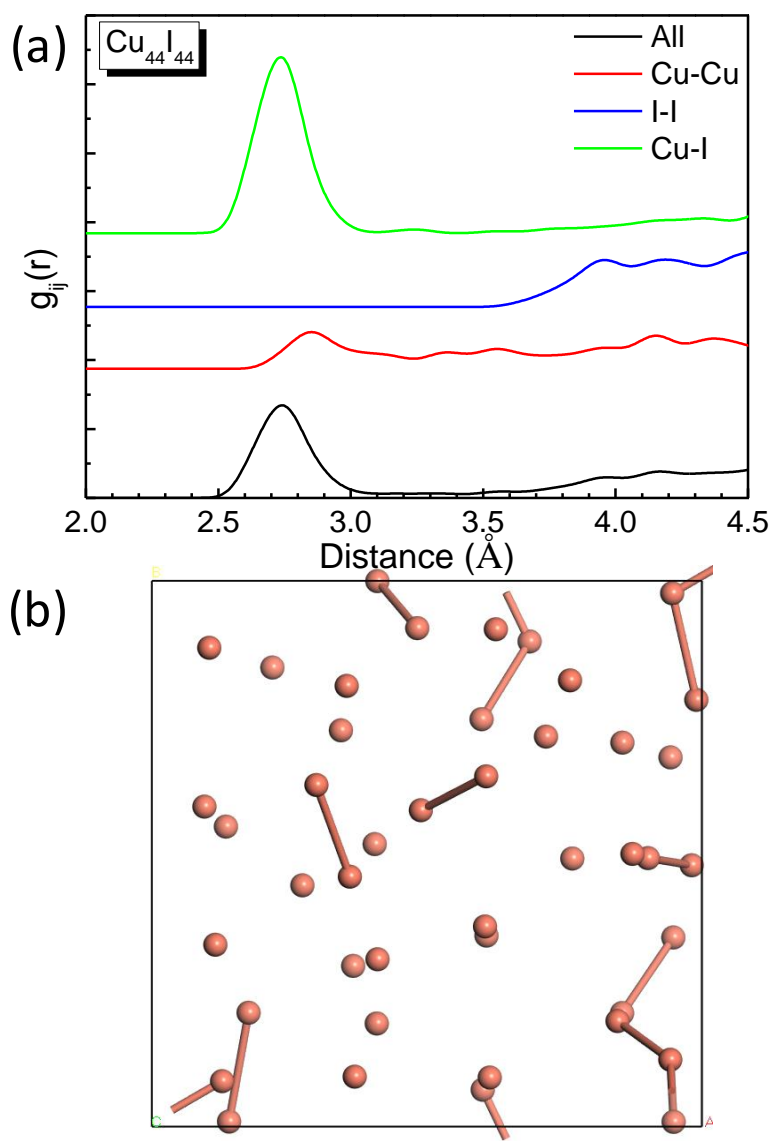

Figure 3. Partial radial distribution functions (RDF) of a-$Cu_{44}I_{44}$ network of Fig. 2.. (b) Network of Fig. 2 showing the Cu-Cu connections with cut-off length of 3.0 Å, indicating the negligible presence of Cu-Cu clusters.

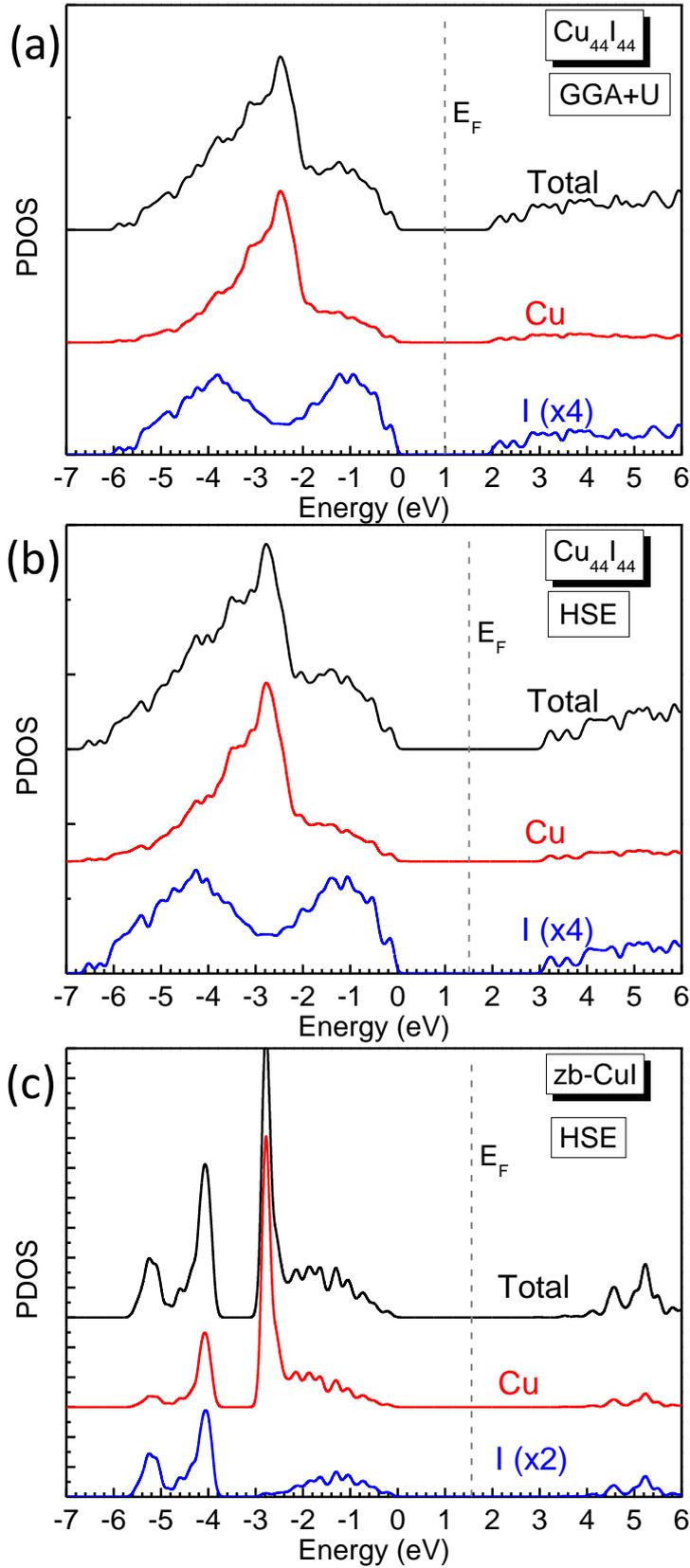

Figure 4. Partial density of states of a-Cu$_{44}$I$_{44}$ network of Fig. 2, calculated by (a) GGA+U and (b) HSE hybrid functional. (c) shows the PDOS of zb-CuI on the same scale for comparison. Note E$_F$ in midgap. The PDOS of I is enlarged to make it more obvious.

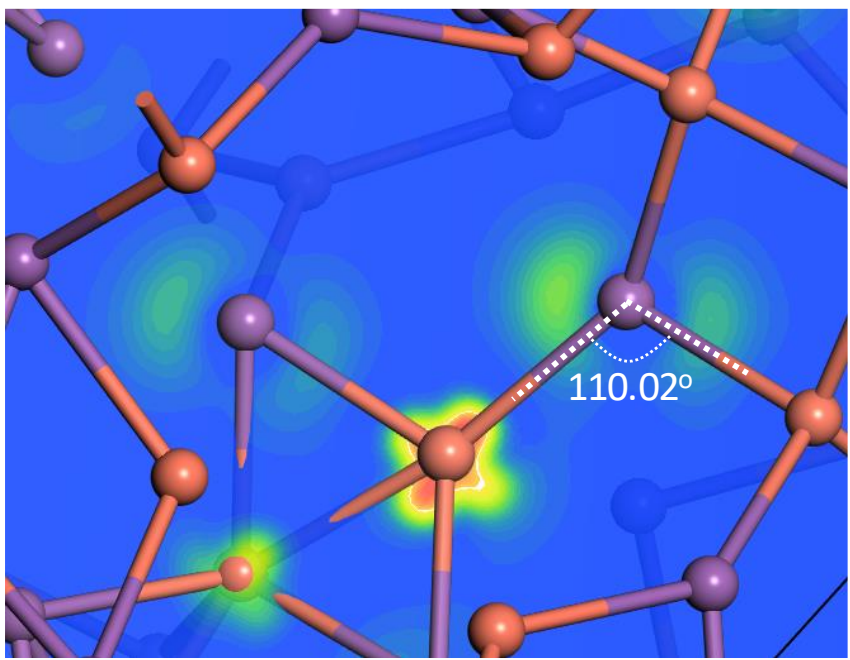

Figure 5. Wavefunction of valence band maximum state in a-Cu$_{44}$I$_{44}$ network of Fig. 2.

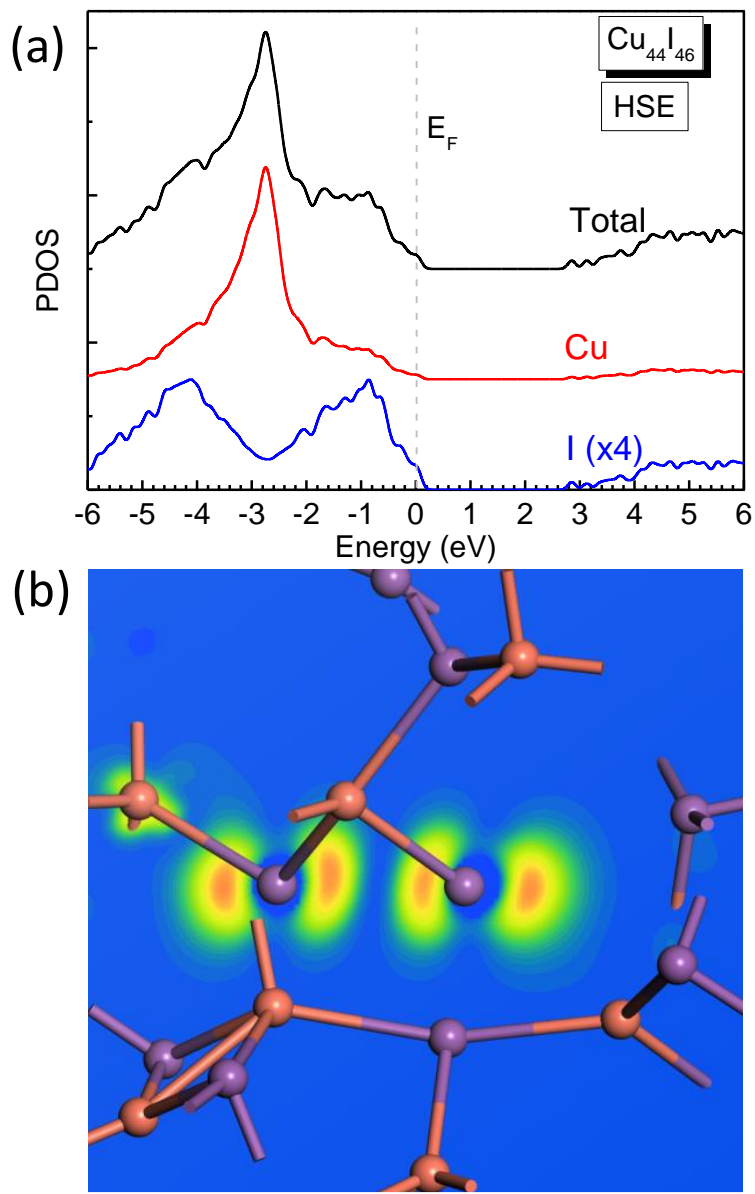

Figure 6. (a) Partial density of states of the I-excess a-$Cu_{44}I_{46}$ network. Note $E_F$ lies in upper valence band. (b) Wavefunction of defect state at top of valence band in I-excess network. Wavefunction localized on two iodine sites.

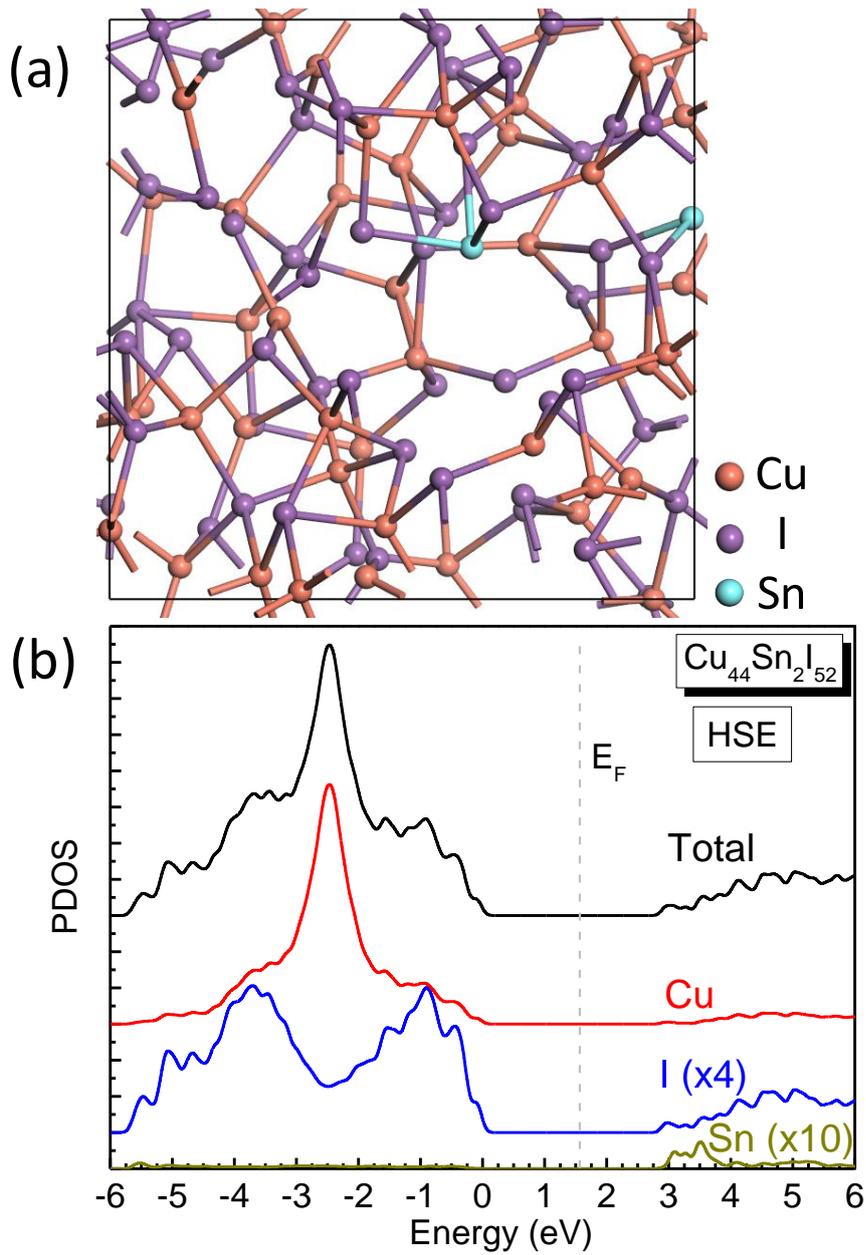

Figure 7. (a) The network of $Cu_{44}Sn_2I_{52}$ with Sn as amorphising agent. Note the greater number of 3-fold I sites, making a more layered structure. (b) PDOS and of this $Cu_{44}Sn_2I_{52}$ network. Note $E_F$ in midgap, as the case of Fig. 4.